\documentclass[reprint,
 amsmath,amssymb,
 aps,
]{revtex4-2}

\usepackage{graphicx}
\usepackage{dcolumn}
\usepackage{bm}
\usepackage{xcolor}
\usepackage{times}
\DeclareGraphicsExtensions{.pdf,.eps,.png,.jpg,.mps} 
\usepackage{ulem}

\newcommand{\FP}{Fabry–P\'erot }

\newcommand{\SN}{Si$_3$N$_4$ }

\begin{document}
 
\title{Spiral resonator referenced on-chip low noise microwave generation }

\author{Long Cheng$^{1*}$, Mengdi Zhao$^{1*}$, Yang He$^{1*}$, Yu Zhang$^{2}$, Roy Meade$^{2}$, Kerry Vahala$^{3}$, Mian Zhang$^{2}$, Jiang Li$^{1}\dagger$\\
$^1$hQphotonics Inc, 2500 E Colorado Blvd Suite 330, Pasadena CA, 91107, USA\\
$^2$HyperLight Corporation, 1 Bow Street, Suite 420, Cambridge, MA 02138, USA\\
$^3$T. J. Watson Laboratory of Applied Physics, California Institute of Technology, Pasadena, CA 91125, USA\\
$^\dagger$Corresponding author: jiang.li@hqphotonics.net
\\
$^*$These authors contribute equally to this work.}

\begin{abstract}

In recent years, miniaturization and integration of photonic microwave oscillators by optical frequency division approach have witnessed rapid progress.   In this work, we report on-chip low phase noise photonic microwave generation  based on a planar chip design.   Dual lasers are co-locked to a silicon nitride spiral resonator and their relative phase noise is measured below the cavity thermal noise limit, resulting in record low  on-chip relative optical phase noise. A broadband integrated electro-optic comb up to 3.43 THz (27 nm) bandwidth is utilized to divide down the relative phase noise of the spiral resonator referenced lasers to the microwave domain. All-around record-low phase noise is achieved for planar chip-based photonic microwave oscillators from 10 Hz to 10 kHz offsets. The planar chip design, high technology-readiness level, foundry-ready processing, combined with the exceptional phase noise performance for our work represent a major advance of integrated photonic microwave oscillators.

\end{abstract}

\maketitle

\section{Introduction}

 Over the last decade, photonic microwave generation based on optical frequency division \cite{Diddams2020} has become the preeminent approach for high performance, ultra-low phase noise microwave generation \cite{fortier2011generation,xie2017photonic,Li2023small}, exceeding the phase noise performance of the best-available electrical oscillators. There has also been rapid progress in miniaturization and integration of photonic microwave oscillators (PMOs) using two-point optical frequency division (2-point OFD)  \cite{li2014electro,sun2024integrated,kudelin2024photonic,zhao2024all,he2024chip}.
 The overall stability of the microwave signals produced by these systems is determined by the dual laser references. And various chip-scale or miniature optical reference cavities, including ultra-high-Q spiral resonators (Q $\sim$ 10$^7$ to 10$^8$ ) \cite{liu202236, he2024chip, sun2024integrated, sun2024kerr}, miniature \FP (FP) cavities (Q $>$ 10$^9$) \cite{kudelin2024photonic,ji2024dispersive}, and discrete  MgF$_2$ crystalline resonators (Q $>$ 10$^9$) \cite{jin2024micro} have been used in these systems.

\begin{figure*}[ht]
\centering
\includegraphics[width=0.8\linewidth]{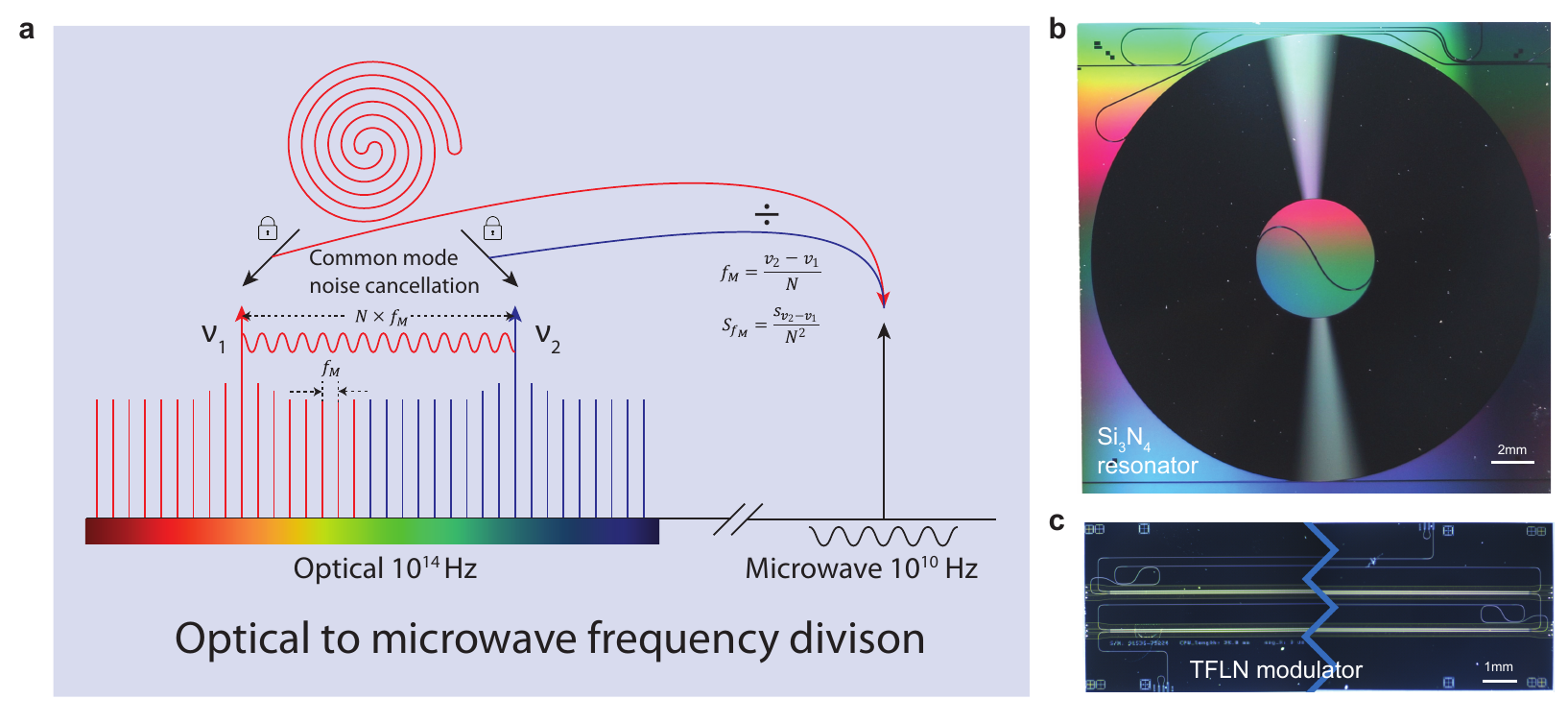}
\caption{  (a) Spiral resonator referenced on-chip low noise
microwave generation architecture based on integrated optical frequency division. The lasers are locked to two cavity modes of an ultra-high Q spiral resonator. Due to common mode noise cancellation of the co-locked lasers from a single resonator,  their frequency difference ($\nu_2 - \nu_1$) achieves low frequency noise below the cavity thermal noise limit. By anchoring the spectral end points of an integrated electro-optic (EO) comb to the dual laser reference, the fractional stability of the dual laser reference  ($\nu_2 - \nu_1$) is  transferred to the EO comb line spacing (microwave rate). And the phase noise of the EO comb line spacing is divided down from the phase noise of the dual laser reference.     (b) Photograph of the 14m-long \SN spiral reference resonator. (c) Image of the tandem thin-film LiNbO$_3$ phase modulator (PM) including two PMs on the same chip. The middle section is not shown. }
\label{fig1}
\end{figure*}

An important feature of 2-point OFD systems is inherent common-mode noise cancellation of two lasers co-locked to a single high-Q resonator \cite{sun2024integrated,kudelin2024photonic,liu2024low}, or by co-lasing such as with dual Brillouin lasers \cite{li2013microwave,gundavarapu2019sub, li2014electro} or optical parametric oscillation \cite{zhao2024all}. Recently, co-PDH locking to a discrete MgF$_2$ resonator \cite{jin2024micro} and a miniature \FP cavity \cite{groman2024photonic} has reduced relative laser phase noise below the absolute cavity thermorefractive noise limit, thereby producing very-low microwave phase noise levels. In this work, we report a significant advancement of chip-based photonic microwave oscillators. By locking two lasers to an ultra-high-Q silicon nitride spiral resonator with Q $>$ 200 million and cavity length of 14 meters, we achieve relative laser noise suppression below the cavity thermorefractive noise (TRN)  limit, achieving a 20 dB improvement in close-in laser phase noise (at 10 Hz offset) compared to other silicon-chip based laser references. This advance, when combined with a wide-span thin-film lithium niobate (TFLN) electro-optic-comb (EO-comb), enables all-around record-low phase noise performance for planar chip-based photonic microwave oscillators from 10 Hz to 10 kHz offsets. Significantly, the planar-chip based PMO demonstrated in this work features a high technology readiness level, foundry-ready processing and low cost manufacturing, which are important for mass-scale applications  of chip-based PMOs in real-world applications including coherent communications,  airborne radar, autonomous driving, and signal processing.

\section{Results}

Figure \ref{fig1}a shows the PMO architecture. Overall frequency stability derives from two lasers co-locked to two cavity modes (frequencies $\nu_1$, $\nu_2$) of a 14-meter-long \SN spiral reference resonator by the Pound-Drever-Hall (PDH) locking technique \cite{drever1983laser}. This stability is transferred to the microwave signal by electro-optical frequency division (e-OFD) using integrated EO-combs.
The absolute frequency noise of each locked laser contains the reference cavity thermorefractive noise (correlated and common-mode noise among the two lasers) and the residual laser locking noise (uncorrelated noise between the two lasers). As demonstrated here, their frequency difference ($\nu_2 - \nu_1$) exhibits frequency noise below the cavity TRN noise limit. 

Figure 1b is a photograph of the Si$_3$N$_4$ spiral resonator. It has a footprint of 21x21 mm and round trip length of 14 meters. The waveguide cross-sectional dimension is 10 um x 0.1 um for low optical confinement and ultra-low on-chip propagation loss ($<$ 0.2 dB/m) at C band wavelengths \cite{jin2021hertz,li2021reaching}. The device is fabricated at a CMOS foundry. It has a loaded quality factor (Q) of 160 million and an intrinsic Q of 204 million. The combination of ultra-high Q factor and large mode volume of the spiral resonator suppresses the laser locking noise as well as the TRN noise of the resonator.  Figure 1c is a photograph of the TFLN chip with tandem phase modulators (PMs). To reduce the V$_\pi$ of the PMs, a  dual-pass (recycled) design reduces the V$_{\pi}$ by a factor two at a given modulation frequency \cite{he2024chip,yu2022integrated}. Also, the long electrode length (26 mm) enhances the electro-optical modulation efficiency \cite{wang2018integrated, kharel2021breaking}.

To establish the system's inherent frequency stability, the relative frequency noise of the lasers, when co-PDH locked to the 14m-long spiral resonator, is first characterized. For this measurement, a RIO laser and a tunable external-cavity diode laser (both operating near 1553 nm) are co-PDH locked to the spiral resonator  and then mixed on a photodetector with a 40 GHz bandwidth. The resulting photodetector beat note (around 20 GHz) is amplified and analyzed on a phase noise analyzer. The measured phase noise between the two co-PDH locked lasers is shown in Figure 2a (blue curve), and reaches levels of -96 dBc/Hz at 10 kHz offset and  -22 dBc/Hz at close-in (10 Hz) offset. For comparison, the red curve in Figure 2a gives the phase noise of the two lasers when free running (no PDH lock). Locking reduces phase noise by 42 dB at 10 kHz offset, and 82 dB at 10 Hz offset. Also plotted in Figure 2a is the simulated TRN noise of the 14m spiral resonator (green curve). The relative phase noise of the co-PDH locked dual laser is suppressed by 6-10 dB from 10 Hz to 10 kHz offset, compared with the absolute cavity TRN noise.

Dual laser references are the backbone of two-point OFD systems and eOFD systems, providing the ultimate fractional frequency stability for the optical-to-microwave frequency-divided outputs. Our co-PDH locked lasers have achieved record-low on-chip relative frequency noise for offset frequencies from 10 Hz to 10 kHz, as shown in Figure 2b.  In particular, at 10 Hz offset, the current work represents $>$ 20 dB reduction of phase noise compared with prior work.

\begin{figure*}[ht]
\centering
\includegraphics[width=\linewidth]{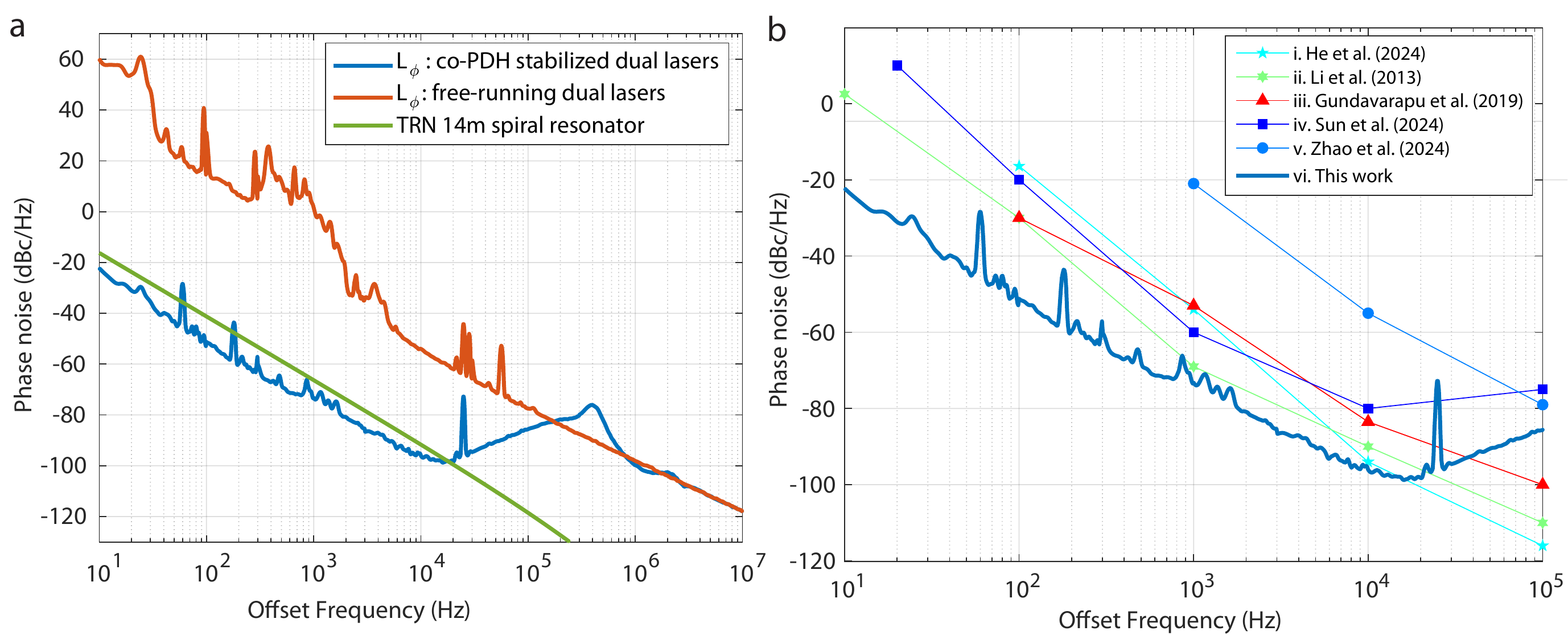}
\caption{ (a) Beat note phase noise of co-PDH locked lasers (blue) and free-running lasers (dashed blue).  Relative phase noise of the co-PDH locked laser to the spiral resonator is reduced below the cavity TRN noise limit (which is shown as the green curve).   (b) Comparison of  silicon-chip-based low noise dual laser references.  The current work has achieved a record low relative frequency noise for on-chip lasers over a broad offset frequency range from 10 Hz to 10 kHz offset.  (b) A comparison of silicon-chip based low noise dual laser references: (i) two lasers co-self-injection-locked to a  \SN spiral resonators \cite{he2024chip}, (ii) dual Brillouin lasers from  a silica disk resonator \cite{li2013microwave}, (iii) dual Brillouin lasers from a \SN ring resonator \cite{gundavarapu2019sub}, (iv) two lasers co-PDH locked to a 4m \SN coil resonator \cite{sun2024integrated},  (v) OPO signal and idler from a \SN ring resonator \cite{zhao2024all},  (vi) this work.  }
\label{fig2}
\end{figure*}

The frequency comb is generated using the tandem TFLN PM chip. The measured Vpi for each individual TFLN PM device is 2.0 V at 37.3 GHz. Figure 3a shows a wide EO comb spectrum with 4.4 THz (35.3 nm) bandwidth (3dB) and 37.3 GHz line spacing generated using this chip where microwave drive power to  each PM device is 35 dBm. Compared with using a single PM device, the tandem PM design doubles the EO comb bandwidth, giving a higher optical-to-microwave frequency division ratio, and 6 dB of extra OFD phase noise reduction. 

Figure 4a gives the schematic for the chip-scale PMO. The two lasers at 1553 nm and 1580 nm are co-PDH locked to the \SN resonator, amplified  to $\sim$100 mW, and coupled to the TFLN PM chip using a lensed fiber. Figure 3b shows the optical spectrum of the co-PDH locked lasers (red) superimposed on  the TFLN EO comb (blue). Two sets of EO combs are shown, each centered around their respective reference laser and having a wavelength span of 27 nm (or 3.43 THz).  After the TFLN chip, the EO combs were amplified using a semiconductor optical amplifier (SOA) to 10 mW, and the central comb spectral lines are bandpass filtered and detected. Here due to the limited tuning range of the optical bandpass filter, we operated the integrated EO comb and the dual laser reference at a narrower span of 3.43 THz for the eOFD operation, instead of the maximum 4.4 THz span achieved in Figure 3a.  The detected signal is used to generate the phase error for feedback control of the VCO via a fast servo filter.

\begin{figure}[ht]
\centering
\includegraphics[width=\linewidth]{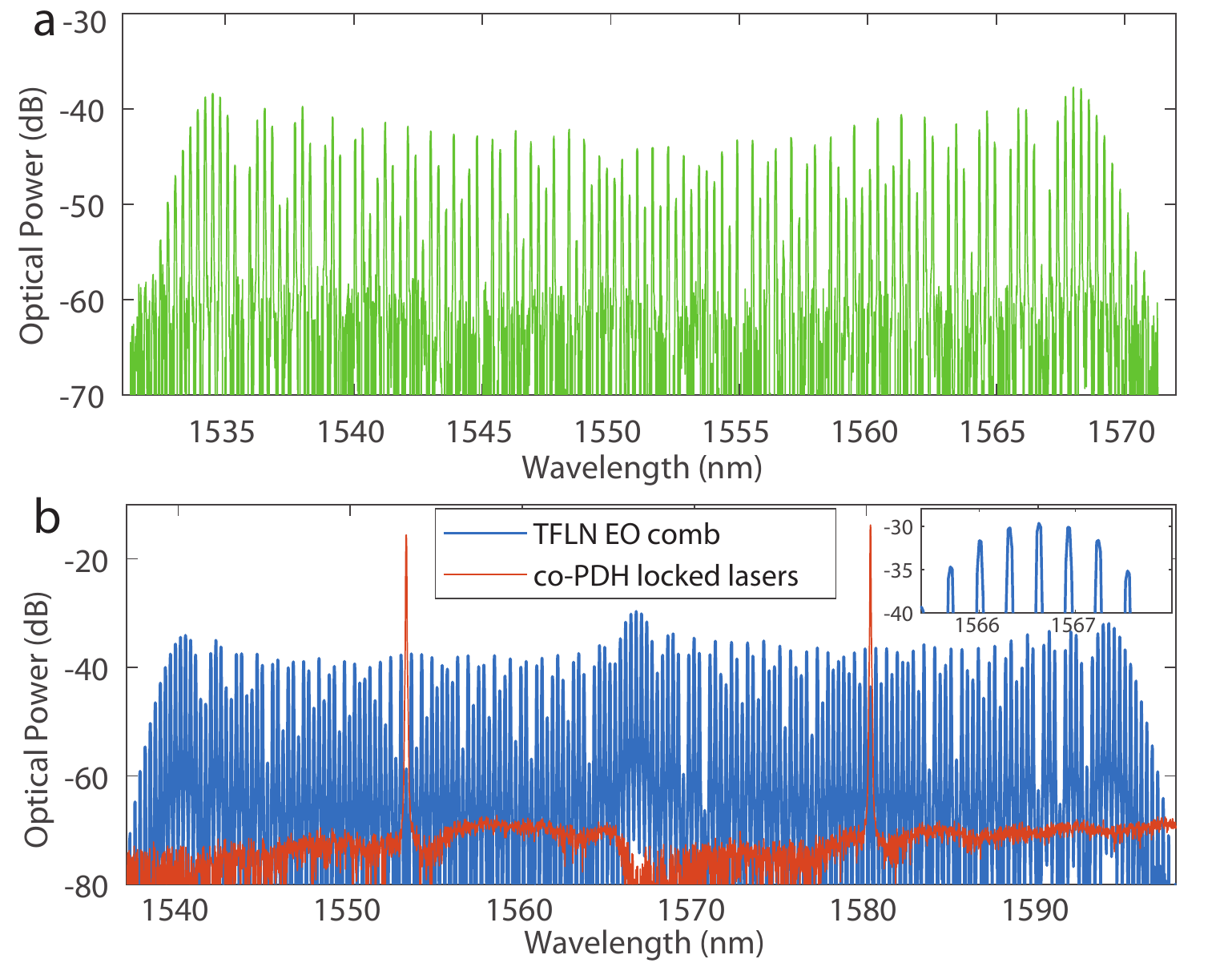}
\caption{ (a). Broadband integrated EO comb spectrum with 3 dB bandwidth of 4.4 THz is generated from the tandem  TFLN PM chip.  (b)   Optical spectra of the co-PDH locked lasers spanning 27 nm (red), and the TFLN EO comb (blue) under eOFD operation.  Inset shows the zoom-in EO comb lines at the spectral middel point between the two lasers.  }
\label{fig3}
\end{figure}

\begin{figure*}[ht]
\centering
\includegraphics[width=0.8\linewidth]{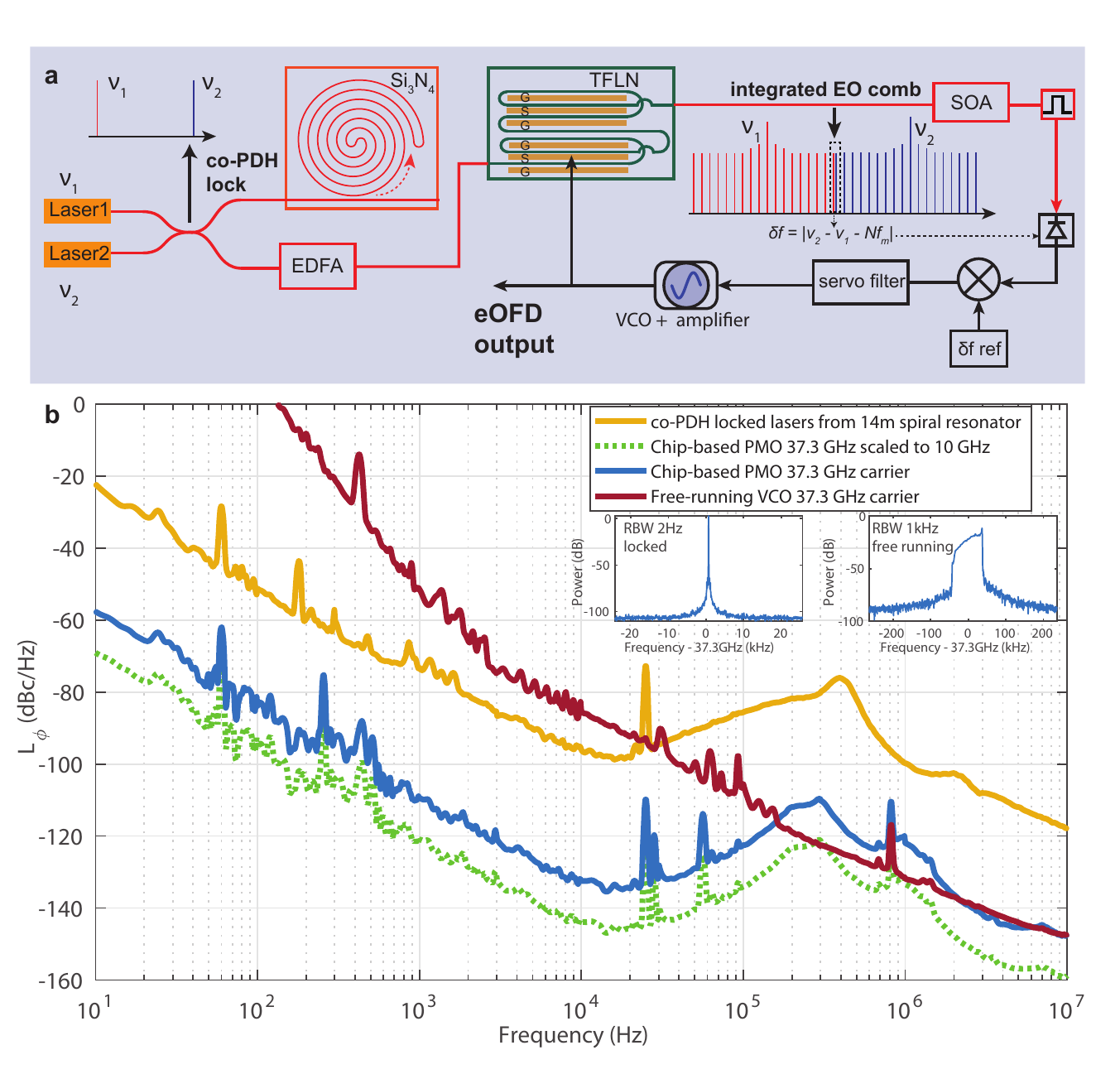}
\caption{ (a) Schematic of the experimental setup for the planar chip-based PMO. (b) Microwave phase noise of the chip-based eOFD oscillator at 37.3 GHz carrier (blue). The dashed green curve is the phase noise of the 10 GHz carrier scaled from the 37.3 GHz eOFD output. Yellow curve is the phase noise of the co-PDH locked lasers. Brown curve is the phase noise of the free running VCO at 37.3 GHz. Insets are the RF spectra for the PMO in locked (RBW 2 Hz) and free-running cases (RBW 1kHz).}
\label{fig4}
\end{figure*}

The optical-to-microwave frequency division ratio (N) for this PMO is 92 (division from 3.43 THz to 37.3 GHz).  \textcolor{black}{The insets in Figure 4b show the RF spectra of the locked oscillator (resolution bandwidth 2 Hz) and the free-running VCO (resolution bandwidth 1 kHz). Clearly the spectral coherence (noise sidebands) is drastically improved (reduced) for the locked case compared with the free-running case.} Figure 4b main panel shows the measured oscillator phase noise at 37.3 GHz (blue curve), along with the free-running VCO phase noise at 37.3 GHz (red curve) and the dual-reference laser phase noise (yellow). Phase noise levels of -133 dBc/Hz at a 10 kHz offset, and -58 dBc/Hz at 10 Hz offset are obtained for the 37.3 GHz carrier (blue curve). The optical-to-microwave frequency division reduces the phase noise of the co-PDH locked lasers uniformly by approximately 39 dB (20log$_{10}$N) from 10 Hz to 10 kHz offset. The dashed green curve gives the phase noise scaled to a 10 GHz carrier (reduction by 20 log$_{10}$[37.3 GHz/10 GHz] dB) for an equivalent phase noise of -144 dBc/Hz at a 10 kHz offset, and -69 dBc/Hz at 10 Hz offset. \textcolor{black}{Of particular note in these spectra are the close-in (10 Hz  offset) phase noise levels which surpass recent silicon-chip-based PMO results by over 10 dB \cite{he2024chip,sun2024integrated,zhao2024all, sun2024kerr}.  Also, at  10 kHz offset (scaled to 10 GHz carrier), the current demonstration surpasses the recent silicon chip-based PMO results by 16 dB \cite{zhao2024all}, 9 dB \cite{sun2024integrated}, 3 dB \cite{he2024chip}, and 2 dB \cite{sun2024kerr}}.

 \section{Conclusion}

In summary, we have demonstrated a low phase noise PMO based on  planar photonic chip design  and integrated electro-optical frequency division.  The device uses high performance planar photonic chips that include an ultra-high-Q \SN spiral resonator and a low V$_{\pi}$ TFLN phase modulator.  The PMO achieves an all-round record-low  phase noise level that greatly improves upon recent silicon-chip-based PMOs.  The exceptional phase noise performance, combined with a planar photonic-chip design, and a high technology readiness level represents a major advance for integrated PMOs. Our demonstration paves the way for chip-based PMOs in mass-scale applications including coherent communications, airborne radar, instrumentation, signal processing, and autonomous driving.

\medskip

\noindent\textbf{Acknowledgments.} This work was supported by the Defense Advanced Research Projects Agency (DARPA)  under contract no. HR001122C0019. The views, opinions and/or findings expressed are those of the authors and should not be interpreted as representing the official views or policies of the Department of Defense or the U.S. Government. Distribution Statement "A" (Approved for Public Release, Distribution Unlimited).
\medskip

\noindent\textbf{Disclosures.} The authors declare no conflicts of interest.  
\medskip



\bibliography{scibib}

\end{document}